\documentclass[conference]{IEEEtran}
\ifCLASSINFOpdf
\else
\fi
\usepackage{amsthm}
\usepackage{amsmath}
\usepackage{amssymb}
\usepackage{graphicx}
\usepackage{balance}
\usepackage{epstopdf}
\usepackage{units}
\usepackage[nospace]{cite}
\usepackage{mathrsfs}
\usepackage{algorithm}

\graphicspath{{Figures/}}

\newcommand*{\E}[1]{\mathsf{E} \left\{ #1 \right\}}

\def \T{\mathsf{T}}
\def \H{\mathsf{H}}
\usepackage{relsize}

\begin{document}


\title{A Quantum of Learning: Using Quaternion Algebra to Model Learning on Quantum Devices}


\author{\IEEEauthorblockN{Sayed Pouria Talebi$^{\dagger}$, Clive Cheong Took$^{\ddagger}$, and Danilo P. Mandic$^{*}$}
\IEEEauthorblockA{$^{\dagger}$Department of Computing, Roehampton University\\$^{\ddagger}$Department of Electronic Engineering, Royal Holloway, University of London\\$^{*}$Electrical \& Electronic Engineering Department, Imperial College London\\
E-mails: sayed.talebi@roehampton.ac.uk, clive.cheongtook@rhul.ac.uk, d.mandic@imperial.ac.uk}}

\maketitle

\begin{abstract}
This article considers the problem of designing adaption and optimisation techniques for training quantum learning machines. To this end, the division algebra of quaternions is used to derive an effective model for representing computation and measurement operations on qubits. In turn, the derived model, serves as the foundation for formulating an adaptive learning problem on principal quantum learning units, thereby establishing quantum information processing units akin to that of neurons in classical approaches. Then, leveraging the modern $\mathbb{HR}$-calculus, a comprehensive training framework for learning on quantum machines is developed. The quaternion-valued model accommodates mathematical tractability and establishment of performance criteria, such as convergence conditions.  
\end{abstract}

\IEEEpeerreviewmaketitle

\section{Introduction}


From the early conceptualisation of neuron activity in the $1940$s~\cite{MP} and initial formulations of  adaptation via neural learning devices in the late $1960$s~\cite{LMS,Net,MP,DaniloBook}, the steady march of computational devices towards increasing capability and decreasing price, has resulted in learning machines becoming intimately woven into our daily lives. This has put research on learning machines and their mechanisms at the frontier of scientific endeavours. Although rapid theoretical advances have been made, machine learning techniques remain computationally demanding. Projecting current demands into the future, considering that Moore's law is beginning to show signs of fatigue~\cite{FromTQ}, and the promise that quantum computers seem to hold~\cite{RF}, suggest that the future of learning machines lie firmly within the quantum realm~\cite{RF}. Reflecting on his experience, conducting simulations of quantum phenomena during the Manhattan project, Richard Feynman pointed out that simulations of quantum mechanics using classical computers appear to scale exponentially in system size and run time~\cite{RF}. This is mainly due to the statistical nature of the quantum realm and is akin to the situation encountered in statistical filtering and learning problems~\cite{Particle,SimonRev}. As a solution to this problem, Feynman surmised whether it would be possible to design a universal computing machine that takes advantage of the statistical nature of quantum phenomena to represent statistical calculations. This has opened a new field focused on deriving information processing techniques based on mathematical operations that can be represented via behaviour of quantum phenomena. 

Although subject of much intrigue among the signal processing~\cite{QuantumSignal,QuantumFiltering}, control~\cite{QuantumControl}, and machine learning~\cite{RevPaper,QL1,QL2} communities, quantum information processing techniques are generally seen to be in their infancy~\cite{Nature}. However, the great potential that quantum computers have been theorised to hold has kindled the interest of researchers in deriving machine learning techniques based on operation of quantum computers~\cite{RevPaper,QuantumControl,QuantumSignal,TheBook,QuantumLecNotes}.  The interest in quantum learning machines has seen efforts to generalise neural network structures to their quantum formulation~\cite{QuantumSignal,Verdon,Nature}. A comprehensive review of these approaches can be found  in~\cite{Nature,RevPaper}. Despite the body of literature that has been dedicated to this field, when it comes to quantum learning machines designed to operate on quantum data, a series of hurdles remain. Theses are; i) formulating suitable quantum generalisation of the neuron as the fundamental learning unit, ii) formulating a suitable optimisation technique for adaptation and learning tasks.  One of the main stumbling blocks in this front seems to be the two-dimensional complex-valued algebra which is used to describe quantum computations. Indeed, an approach that seems to be gaining traction is the use of quaternions to model qubits and their computations~\cite{Kliuchnikov2015a,Kliuchnikov2015AFF,Me,10558741,10558742}. The use of higher order algebras, such as quaternions, in quantum computing initiated at Microsoft~\cite{Kliuchnikov2015a,Kliuchnikov2015AFF} while investigating solutions to the compiler problem. Furthermore, the use of quaternions has recently been shown advantageous in processing images via quantum computers~\cite{10558741,10558742}. However, to our best knowledge, use of the quaternion algebra for deriving a learning structures for processing quantum data remains unaddressed.   

This article focuses on effective use of quaternions and their associated $\mathbb{HR}$-calculus to derive fundamental adaptation and learning solutions for quantum information processing machines. To this end, quantum computations on qubits are modelled using the division algebra of quaternions and accessible information via measurement operations on qubits is derived. Subsequently, the fundamental learning problem on quantum data is formulated and a mathematically tractable solution is provided.

\noindent\textbf{\textit{Nomenclature}}: Scalars, column vectors, and matrices are denoted by lowercase, bold lowercase, and bold uppercase letters, whereas $\mathbb{N}$, $\mathbb{R}$, $\mathbb{C}$, and $\mathbb{H}$ denote set of natural, real, complex, and quaternion numbers, while remainder of the nomenclature is summarised as follows:

\begin{tabular}{ll}
$\{\imath,\jmath,\kappa\}$ & quaternion imaginary units
\\
$\wedge$ & logical conjunction 
\\
$\mathsf{P}\left(\cdot\right)$ & probability density function
\\
$\E{\cdot}$ & statistical expectation operator
\\
$(\cdot)^{\T}$ - $\left(\cdot\right)^{\H}$ & transpose - Hermitian transpose operators
\\
 $\mathbf{I}$ & identity matrix of appropriate size
\\
$\left\lceil\cdot \right\rceil$ & ceiling operator 
\end{tabular}

\section{Quaternion Algebra}

quaternions are a  four-dimensional skew-field so that a quaternion variable $q \in \mathbb{H}$ consists of a real part, $\Re\{q\}$, and a three-dimensional imaginary part $\Im\{q\}$, comprised of the components, $\Im_{\imath}\{q\}$, $\Im_{\jmath}\{q\}$, and $\Im_{\kappa}\{q\}$. Hence, $q$ can be expressed as $q =\Re\{q\} + \Im\{q\}$ or alternatively 
\[
q= \Re\{q\} + \Im_{i}\{q\} + \Im_{j}\{q\} + \Im_{k}\{q\}=q_{r} + \imath q_{\imath} + \jmath q_{\jmath} + \kappa q_{\kappa}
\]
where $\{q_{r},q_{\imath},q_{\jmath},q_{\kappa}\}\subset\mathbb{R}$, while $\imath$, $\jmath$, and $\kappa$ are imaginary units obeying the following product rules
\begin{equation}
\imath\jmath=\kappa,\jmath\kappa=\imath,\kappa\imath=\jmath,\imath^{2}=\jmath^{2}=\kappa^{2} =\imath\jmath\kappa=-1.
\label{eq:Multiplication}
\end{equation}
Most notably, it follows from \eqref{eq:Multiplication} that multiplication is not a commutative affair in the quaternion domain.\footnote{For example, form \eqref{eq:Multiplication} note that $\jmath=\kappa\imath$, and thus, $\jmath\imath=\kappa\imath\imath=-\kappa\neq\imath\jmath$.} The conjugate and norm of $q\in\mathbb{H}$ are respectively given by 
\[
q^{*}=\Re\{q\}-\Im\{q\}\hspace{0.12cm}\text{and}\hspace{0.12cm}\|q\|=\sqrt{qq^{*}}.
\]
A quaternion $q\in\mathbb{H}$ can alternatively be expressed in its polar presentation, given by~\cite{Me}
\[
q =\|q\|e^{\xi \theta}=\|q\|\big(\text{cos}(\theta)+\xi\text{sin}(\theta)\big)
\]
where $\xi = \frac{\Im\{q\}}{\|\Im\{q\}\|}$ and $\theta=\text{atan}\left(\frac{\|\Im\{q\}\|}{\Re\{q\}}\right)$. 

\begin{figure}[h]
\centering
\includegraphics[width=1\linewidth,trim = 0cm 0cm 0cm 0cm]{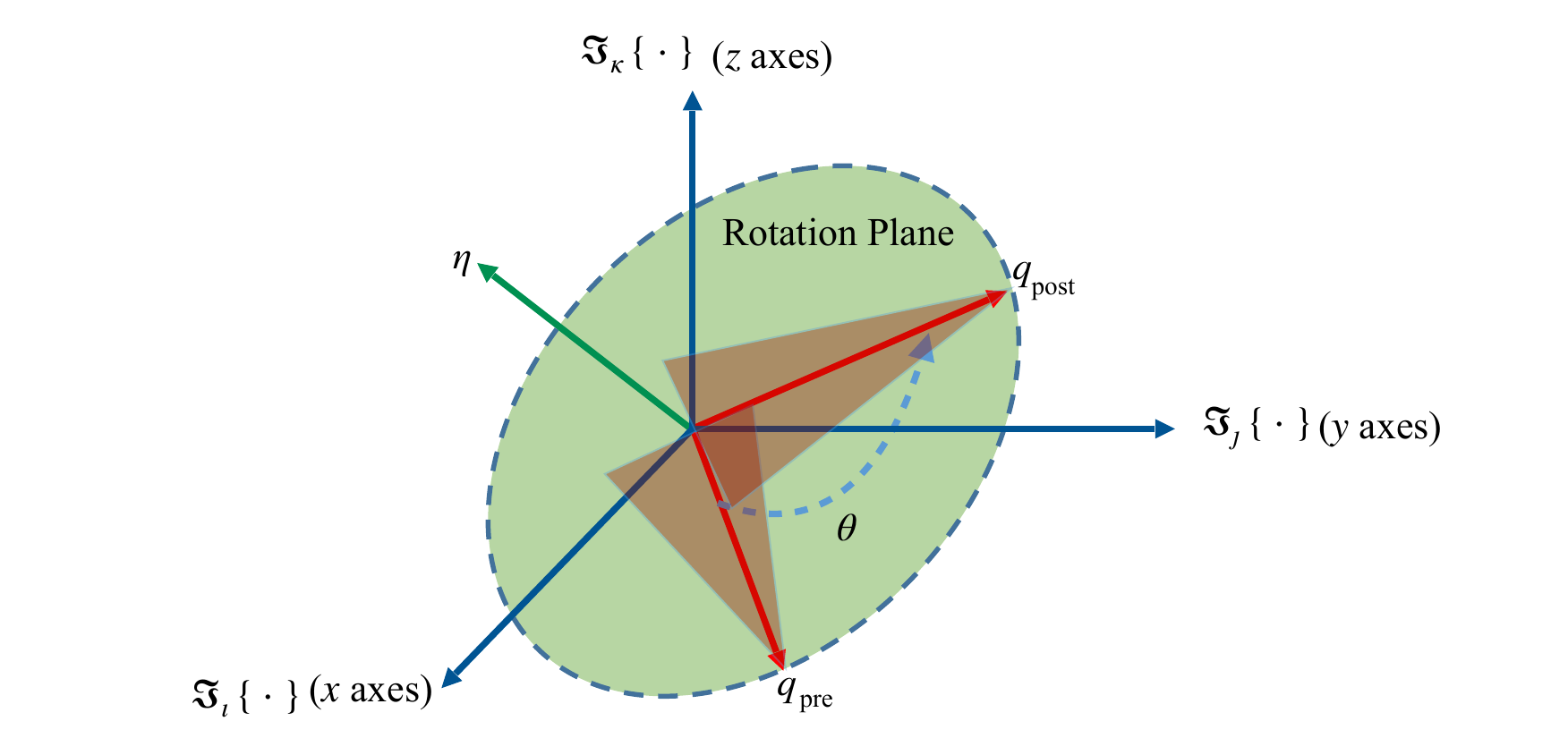}
\caption{Schematic showing the right-hand rotation of $q_{\text{pre}}$ around $\eta$ by an angle of $\theta$ to arrive at $q_{\text{post}}$.}
\label{fig:Rotation}
\end{figure}

In the quaternion domain, involution of $q\in\mathbb{H}$ around $\zeta\in\mathbb{H}\setminus0$ is defined as $q^{\zeta} = \zeta q \zeta^{-1}$~\cite{InvoQ,CliveHR}. The traditional application of quaternion involutions has been in modelling rotations in three-dimensional space. To this end, consider the Cartesian coordinates of an object prior to rotation to be given by $q_{\text{pre}}=(x,y,z)$ and modelled as the quaternion $q_{\text{pre}}=\imath x+\jmath y+\kappa y$. Then, the right-hand rotation of $q_{\text{pre}}$ around the unit vector $\eta$ by the angle $\theta$, as shown in Fig.~\ref{fig:Rotation}, is expressed in terms of quaternion involutions 
\begin{equation}
q_{\text{post}}=\xi q_{\text{pre}}\xi^{-1}\hspace{0.12cm}\text{with}\hspace{0.12cm}\xi=e^{\eta\frac{\theta}{2}}=\cos\left(\frac{\theta}{2}\right)+\eta\sin\left(\frac{\theta}{2}\right)\cdot
\label{eq:QuaternionRotation}
\end{equation}
More recently, quaternion involutions have been used to establish a one-to-one linear relation between involutions of $q\in\mathbb{H}$ and its real-valued components given by
\begin{equation}
\mathbf{q}^{a}=\begin{bmatrix}\mathbf{q}\\\mathbf{q}^{\imath}\\\mathbf{q}^{\jmath}\\\mathbf{q}^{\kappa}\end{bmatrix}=\begin{bmatrix}\mathbf{I}&\imath\mathbf{I}&\jmath\mathbf{I}&\kappa\mathbf{I}\\\mathbf{I}&\imath\mathbf{I}&-\jmath\mathbf{I}&-\kappa\mathbf{I}\\\mathbf{I}&-\imath\mathbf{I}&\jmath\mathbf{I}&-\kappa\mathbf{I}\\\mathbf{I}&-\imath\mathbf{I}&-\jmath\mathbf{I}&\kappa\mathbf{I}\end{bmatrix}\begin{bmatrix}\mathbf{q}_{r}\\\mathbf{q}_{\imath}\\\mathbf{q}_{\jmath}\\\mathbf{q}_{\kappa}\end{bmatrix}
\label{eq:real-valued components}
\end{equation}
where $\mathbf{q}^{a}$ is referred to as the augment quaternion vector. Importantly, the rotation in \eqref{eq:QuaternionRotation} can be formulated in a linear manner in the augmented quaternion space so that~\cite{Me,MeStatQ}
\begin{equation}
\exists \mathbf{Z}\in\mathbb{H}^{4\times4}:\mathbf{q}^{a}_{\text{post}}=\mathbf{Z}\mathbf{q}^{a}_{\text{pre}}.
\label{eq:AugRot}
\end{equation}
with elements of $\mathbf{Z}$ being tractably related to $\eta$ and $\theta$. 

Using the relation in \eqref{eq:real-valued components},  the $\mathbb{HR}$-calculus~\cite{CliveHR,Me}, considers the function $\mathsf{f}(\mathbf{q}=\mathbf{q}_{r}+\imath\mathbf{q}_{\imath}+\jmath\mathbf{q}_{\jmath}+\kappa\mathbf{q}_{\kappa}):\mathbb{H}^{M}\rightarrow\mathbb{H}$ in terms of the quaternion-valued basis $\{\mathbf{q},\mathbf{q}^{\imath},\mathbf{q}^{\jmath},\mathbf{q}^{\kappa}\}$, so that we have $\mathsf{f}(\mathbf{q}^{a}=[\mathbf{q}^{\T},\mathbf{q}^{\imath\T},\mathbf{q}^{\jmath\T},\mathbf{q}^{\kappa\T}]^{\T}):\mathbb{H}^{4M}\rightarrow\mathbb{H}$. Expressing the quaternion-valued function in terms of its real-valued components as $\mathsf{f}(\mathbf{q}^{a})=\mathsf{f}_{r}(\mathbf{q}^{a})+\imath\mathsf{f}_{\imath}(\mathbf{q}^{a})+\jmath\mathsf{f}_{\jmath}(\mathbf{q}^{a})+\kappa\mathsf{f}_{k}(\mathbf{q}^{a})$ and using the mapping in (\ref{eq:real-valued components}), a relation is established between the derivatives taken in $\mathbb{R}^{4}$ and those taken directly in $\mathbb{H}$. This, forms a unified framework for calculating the derivatives and establishing the gradients of quaternion-valued functions directly in the quaternion domain~\cite{Me,CliveHR}.

\noindent\textbf{Remark~1:} The $\mathbb{HR}$-calculus establishes the direction of steepest change for $\mathsf{f}\left(\mathbf{q}^{a}\right)$ as the derivative $\frac{\partial\mathsf{f}\left(\mathbf{q}^{a}\right)}{\partial\mathbf{q}^{a*}}$~\cite{PouriaPHD,Me,CliveHR}. 

\section{Formulating Quantum Computation with Quaternion Algebra}

In quantum computing, the basic unit of information is the qubit. The qubit exists in a superposition of two orthogonal bases denoted with $|0\rangle$ and $|1\rangle$. These units represent state of a quantum particle~\cite{QuantumControl}. The qubit can be in any superposition that satisfies
\begin{equation}
\forall\{\alpha,\beta\}\subset\mathbb{C}\wedge\|\alpha\|^{2}+\|\beta\|^{2}=1:\hspace{0.12cm}|\psi\rangle=\alpha|0\rangle+\beta|1\rangle
\label{eq:QubitDef}
\end{equation}
where $|\psi\rangle$ represents state of the qubit while $\|\alpha\|^{2}$ and $\|\beta\|^{2}$ correspond to the probability of measuring the quantum particle in state $|0\rangle$ or $|1\rangle$. 

\noindent\textbf{Remark~2:} Note that the state of $|\psi\rangle$ is considered unchanged if $\alpha$ and $\beta$ are rotated by the same angle, that is if $\alpha$ and $\beta$ are multiplied by $e^{\imath\varphi}$ with $\varphi\in\mathbb{R}$. 

\begin{figure}[h]
\centering
\includegraphics[width=1\linewidth, trim = 0cm 2.8cm 0cm 0cm]{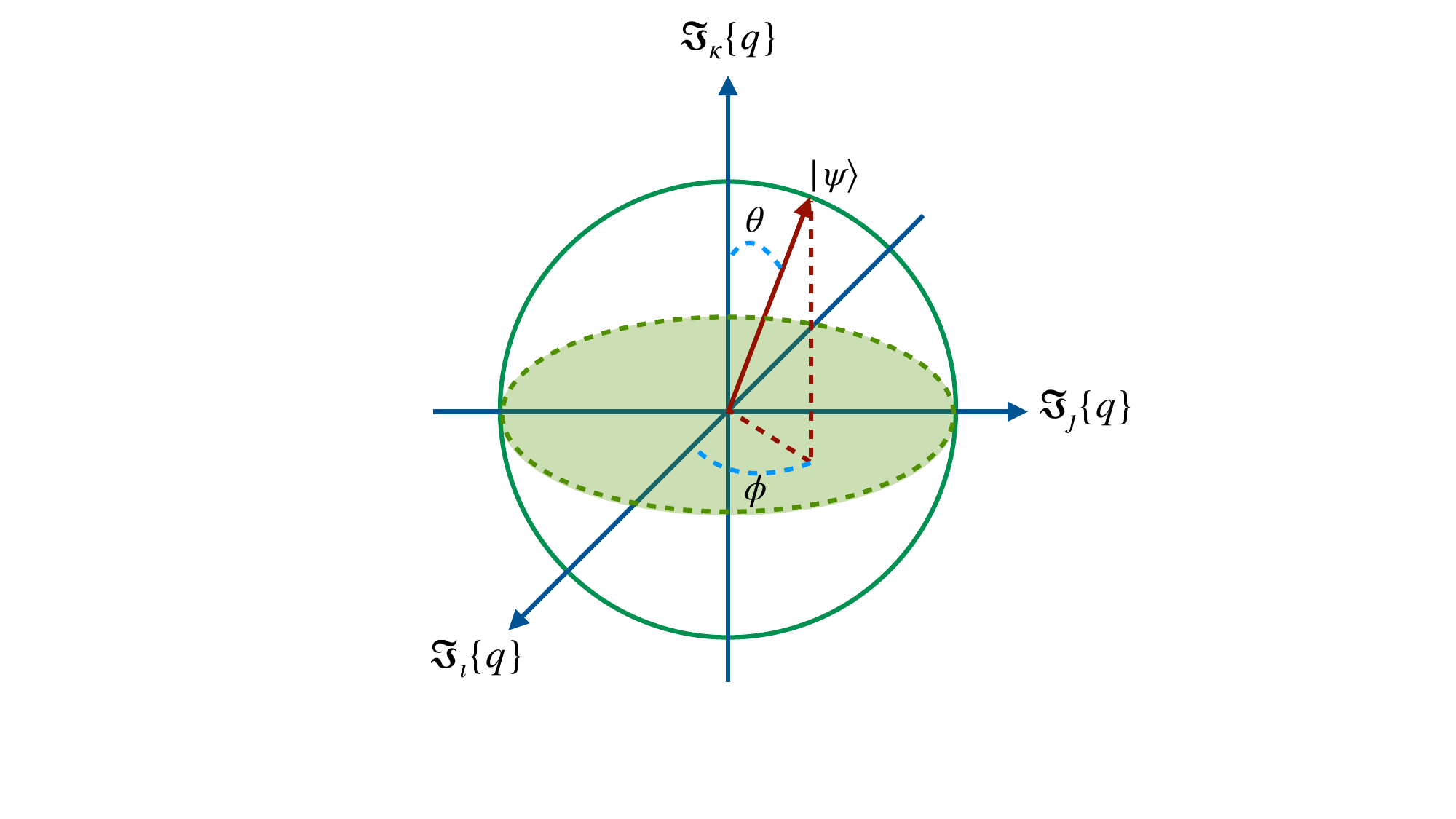}
\caption{Bloch sphere representation of qubit $|\psi\rangle$ modelled as the quaternion $q=\imath\sin(\theta)\cos(\phi)+\jmath\sin(\theta)\sin(\phi)+ \kappa\cos(\theta)$. }
\label{fig:QuatQuant}
\end{figure}

Although the two complex-valued variables $\alpha$ and $\beta$ might suggest four degrees of freedom for $|\psi\rangle$, the unit sum condition in \eqref{eq:QubitDef} and \textit{Remark}~2, limit the degrees of freedom to two. Thus, the state of $|\psi\rangle$ can be described via a pure unit imaginary quaternion as demonstrated in Fig.~\ref{fig:QuatQuant} where  
\begin{equation}
|\psi\rangle\leftrightarrow q=\imath\sin(\theta)\cos(\phi)+\jmath\sin(\theta)\sin(\phi)+ \kappa\cos(\theta).
\end{equation} 
The presentation of a qubit on the unit three-dimensional sphere is referred to as the Bloch sphere representation~\cite{TheBook,QuantumControl}. 

\noindent\textbf{Remark~3:} An elegant solution for repressuring the state of $|\psi\rangle$ in the quaternion domain is to rotate $\alpha$ and $\beta$ components of $|\psi\rangle$ so that $\Im\{\alpha\}\rightarrow0$; then, use $\alpha$, $\Re\{\beta\}$, and $\Im\{\beta\}$ as $\kappa$, $\imath$, and $\jmath$ components of the quaternion. Due to \textit{Remark}~2, this does not effect the state of qubit being represented. 


\noindent\textbf{Proposition~1:} All quantum computing gates represent a rotation operation on a qubit, that is a quaternion involution, on the quaternion-valued model of the qubit. In addition, computations on multi-qubit system are representable via linear unitary mappings in the augmented quaternion space.  

\noindent\textit{Proof of Proposition~1:} Consider the qubit $|\psi_{\text{out}}\rangle$ and $|\psi_{\text{in}}\rangle$ related via quantum gate $\mathbf{U}$, so that $|\psi_{\text{out}}\rangle=\mathbf{U}|\psi_{\text{in}}\rangle$, where $\mathbf{U}\in\mathbb{C}^{2\times2}$ is a unitary\footnote{A unitary transform preserves norm of the input state, ensuring that amplitudes of the input and output states remain one.} transform, $\langle\psi_{\text{out}}|\psi_{\text{out}}\rangle=1$, and $\langle\psi_{\text{in}}|\psi_{\text{in}}\rangle=1$ to satisfy the unit sum condition in \eqref{eq:QubitDef}. As a result, quaternions $q_{\text{out}}$ and $q_{\text{in}}$ modelling $|\psi_{\text{out}}\rangle$ and $|\psi_{\text{in}}\rangle$, will be pure imaginary quaternion-valued variables with unit norm. It follows from geometric algebra that $q_{\text{in}}$ and $q_{\text{out}}$ denote a plane in the quaternion imaginary sub-space~\cite{Silverman,FreQ}. In this setting, $\eta=\Im\{q_{\text{in}}q_{\text{out}}\}$ is normal to the plane that contains $q_{\text{out}}$ and $q_{\text{in}}$~\cite{FreQ}. Therefore, if $\theta$ denotes the angle between $q_{\text{out}}$ and $q_{\text{in}}$ in the plane that contains them, the quaternion involution in \eqref{eq:QuaternionRotation} formulates the relation between $q_{\text{in}}$ and $q_{\text{out}}$ modelling the operation of quantum gate $\mathbf{U}$ in the quaternion domain. This operation is linear in the augmented quaternion space\footnote{Also referred to as widely-linear in the literature, see~\cite{Me,MeStatQ}.} as expressed in \eqref{eq:AugRot}. Since in \eqref{eq:AugRot} the input and output have the same amplitude, the transform is unitary. Now, consider an $m$-qubits system, the state of which in representable via $\mathbf{q}=\frac{1}{\sqrt{m}}\begin{bmatrix}q_{1},\ldots,q_{m}\end{bmatrix}^{\T}$, where $q_{i}$ represents state of the $i^{\text{th}}$ qubit. Due to the nature of quantum particles, two class of operations are possible; i) a rotation of each qubit, ii) summation and re-normalisation of subset of qubits or their rotated versions~\cite{QuantumControl,QuantumLecNotes,TheBook}, both of which are linear transformations in the quaternion augmented space~\cite{Me}. Therefore, all possible computations on $m$-qubit systems are linear unitary transforms in the augmented quaternion space.

In an  $m$-qubit system, measurement against an arbitrary state determines whether the system collapses into that state and is a probabilistic process~\cite{QuantumLecNotes,QuantumSignal,TheBook}. This is formalised in the quaternion domain so that for $\mathbf{h}\in\mathbb{H}^{m}$ with $\Re\{\mathbf{h}\}=\mathbf{0}$ and $\|\mathbf{h}\|^{2}=1$, representing a valid quantum state, measurement of $\mathbf{q}$ against $\mathbf{h}$ results in the following probabilities
\begin{equation}
\begin{aligned}
\mathsf{P}\left(\mathbf{q}=\mathbf{h}\right)=&\frac{1}{4}\|\Re\{\mathbf{h}^{a\T}\mathbf{q}^{a}\}\|^{2}
\\
\mathsf{P}\left(\mathbf{q}\neq\mathbf{h}\right)=&\frac{1}{4}\|\Im\{\mathbf{h}^{a\T}\mathbf{q}^{a}\}\|^{2}
\end{aligned}
\label{eq:Measurment}
\end{equation}
This, allows quantum computations to be interpreted as functions on probability distributions. The expressions in \eqref{eq:Measurment} follows from the theory of quantum measurement~\cite{TheBook}. In this setting, consider the set of quaternion-valued vectors $\mathcal{H}=\{\mathbf{h}_{1},\ldots,\mathbf{h}_{m}\}$ that lie within the measurable subspace of an $m$-qubit  system. Then, the measurement of $m$-qubit system alongside $\mathbf{h}_{i}\in\mathcal{H}$ is represented as the statistical outcome
\[
\mathsf{M}_{\mathbf{h}_{i}}\left(\mathbf{q}\right)\rightarrow\left\{\begin{matrix}1\hspace{0.12cm}\text{with probability}\hspace{0.12cm}\frac{1}{4}\|\Re\{\mathbf{h}^{a\T}_{i}\mathbf{q}^{a}\|^{2}\}\\\\0\hspace{0.12cm}\text{with probability}\hspace{0.12cm}\frac{1}{4}\|\Im\{\mathbf{h}^{a\T}_{i}\mathbf{q}^{a}\|^{2}\}\end{matrix}\right.
\]
where post-measurement the $m$-qubit system collapses into $\mathbf{h}_{i}$ if the outcome is $1$ and exists in a superposition of vector basis $\mathcal{H}\setminus\mathbf{h}_{i}$ if the outcome is $0$.\footnote{The statement implicitly assumes that the system in question has not collapsed into definite sate prior to measurements} Next, the derived concepts of observation and quantum computation are used to develop an adaptive learning technique within the quantum domain.

\section{Quantum Learning}

Let us consider the data set $\mathcal{D}=\{\mathcal{D}_{n}:n\in\mathbb{N}\}$ with sub-set $\mathcal{D}_{n}\sim\mathsf{P}\left(\mathcal{D}_{n}|\zeta\right)$ where $\zeta$ denotes  a hypothesis variable. The aim is to find the mapping $\mathsf{f}:\mathsf{P}\left(\mathcal{D}_{n}|\zeta\right)\rightarrow\mathsf{P}\left(\zeta|\mathcal{D}_{n}\right)$, that is, a mapping between probability distribution of a data set  and that of the possible hypothesis generating that data set. This is a class of problems with application in Bayesian learning~\cite{BL}, engineering~\cite{Def}, and sequential filtering~\cite{Particle,SimonRev}. In order to formulate the quantum learning problem, the probability distribution $\mathsf{P}\left(\mathcal{D}_{n}|\zeta\right)$ is considered to be encoded as a multi-qubit system denoted with $\mathbf{x}_{n}$, whereas probability distributions $\mathsf{P}\left(\zeta|\mathcal{D}_{n}\right)$ is encoded and denoted with $\mathbf{y}_{n}$.  

\noindent\textbf{Assumption~1:} It is assumed that $\mathsf{P}\left(\mathcal{D}_{n}|\zeta\right)$ is known, which is in-line with current assumptions made within the statistical machine learning community~\cite{Def,BL,Particle}.

The aim is to estimate the mapping from $\{\mathbf{x}_{n}:n\in\mathbb{N}\}$ to $\{\mathbf{y}_{n}:n\in\mathbb{N}\}$, given the available data set $\mathcal{D}$ using only two sets of operations; i) rotation of one qubit, ii) addition with renormalisation of two or more qubits. This operations are selected as they are implementable via known quantum circuitry~\cite{QuantumControl,QuantumLecNotes} and straightforward to be applied while being both linear in the quaternion augmented framework~\cite{Me} and facilitating structured learning and optimization within quantum machine learning models somewhat akin to the role of neurons in modern artificial neural networks. The problem can now be expressed mathematically as 
\begin{equation}
\min_{\mathbf{W}}:\hspace{0.12cm}\mathsf{J}\left(\mathbf{W}\right)=\E{\frac{1}{m}\sum^{m}_{s=1}\mathsf{d}\left(\hat{d}_{s,n},d_{s,n}\right)}
\label{eq:RealCost}
\end{equation}
where the expectation is taken over the available data set $\mathcal{D}$ with $\mathsf{d}\left(\cdot,\cdot\right)$ denoting a metric function, whereas
\begin{equation}
\hat{d}_{s,n}=\E{\mathsf{M}_{\mathbf{h}_{s}}\left(\mathbf{W}\mathbf{x}^{a}_{n}\right)}\hspace{0.12cm}\text{and}\hspace{0.12cm}d_{s,n}=\E{\mathsf{M}_{\mathbf{h}_{s}}\left(\mathbf{y}^{a}_{n}\right)}
\label{eq:MeasurmentOb}
\end{equation}
with expectations indicate expected measurements outcomes. In essence,  we aim to find the circuit that $\mathbf{W}$ represents and fits desired  measurement outcomes $\{d_{s,n}:s=1,\ldots,m,n\in\mathbb{N}\}$ to their estimates $\{\hat{d}_{s,n}:s=1,\ldots,m,n\in\mathbb{N}\}$.  

\noindent\textbf{Remark~4:} Consider a mode of $\mathbf{x}^{a}_{n}$ that represents the state of one qubit. The multiplication of that mode in $\mathbf{x}^{a}_{n}$ with corresponding sub-elements in $\mathbf{W}$ models a unitary rotation in the state of the qubit in question with other operations in the matrix multiplication representing operations, e.g. conditional rotations, that are synthesisable  via quantum circuits~\cite{QuantumControl,QuantumLecNotes}. 

\noindent\textbf{Theorem~1:} The expectations in \eqref{eq:MeasurmentOb} must be available alongside $3\left\lceil\log_{2}\left(z\right)\right\rceil$ linearly independent vectors for the mapping $\mathsf{f}:\mathsf{P}\left(\mathcal{D}_{n}|\zeta\right)\rightarrow\mathsf{P}\left(\zeta|\mathcal{D}_{n}\right)$ to be recoverable.   
 
\noindent\textit{Proof of Theorem~1:} A qubit can encode the probability of one hypothesis as the amplitude of one of its states (either $|0\rangle$ or $|1\rangle$) with the amplitude of the other state being determined via the unit sum condition form \eqref{eq:QubitDef}. In this setting, $z$ hypothesis are encodable via an $\left\lceil\log_{2}\left(z\right)\right\rceil$-qubit systems. On the other hand, the optimisation problem in \eqref{eq:RealCost} is equivalent to the distributed state estimation problem in~\cite{DistQuatControl} with $m$ agents in a fully connected network, where agent $s$ has access to measurements $d_{s,n}$. From the work in~\cite{DistQuatControl} it follows that the class of problem is \eqref{eq:RealCost} have a solution if the state is measurable along $3\left\lceil\log_{2}\left(z\right)\right\rceil$-independent vectors.\footnote{The reader is referred to Proposition~1~and~2 from~\cite{DistQuatControl}.} 

The condition in \textit{Theorem~1} sets requirement for a solution to \eqref{eq:RealCost} to exist. However, from \eqref{eq:Measurment} note that statistical expectation of quaternion measurements can only reveal the amplitude of $m$-qubit system alongside a possible state. Thus, a qubit cannot be uniquely reconstructed from its measurements and a unique solution to the problem in \eqref{eq:RealCost} cannot be formulated from measurements. For the optimisation problem in \eqref{eq:RealCost} to be solvable a further assumptions are needed. 

\noindent\textbf{Assumption~2:} It is assume $\mathsf{P}\left(\zeta|\mathcal{D}_{n}\right)$ is encoded so that for all $n$ and $s$, we have $\Re\{\mathbf{h}^{a\T}*\mathbf{y}^{a}_{n}\}>0$.

\noindent\textbf{Remark~5:} As encoding are determined via the user , \textit{Assumption}~2, is not a restrictive one. In addition, this criteria can be use to introduce nonlinearity into the learning process somewhat similar to the role of the activation function. 

Solution to the optimization problem in \eqref{eq:RealCost} is obtainable through a gradient descent strategy, where we have
\[
\hat{\mathbf{W}}_{n+1}=\hat{\mathbf{W}}_{n}-\mu\nabla_{\hat{\mathbf{W}}^{*}_{n}}\mathsf{J}\left(\hat{\mathbf{W}}_{n}\right)
\]  
with $\mu$ denoting a real-valued positive adaptation gain and  $\nabla_{\hat{\mathbf{W}}^{*}_{n}}\mathsf{J}\left(\mathbf{W}\right)$ denoting the $\mathbb{HR}$-gradient of $\mathsf{J}\left(\hat{\mathbf{W}}_{n}\right)$ while $\hat{\mathbf{W}}_{n}$ denotes the estimate of $\mathbf{W}$. Using the chain derivative rule of the $\mathbb{HR}$-calculus, we have 
\[
\nabla_{\hat{\mathbf{W}}^{*}_{n}}\mathsf{J}\left(\hat{\mathbf{W}}_{n}\right)=\frac{1}{m}\mathlarger{\sum}^{m}_{s=1}\frac{\partial\mathsf{d}\left(\hat{d}_{s,n},d_{s,n}\right)}{\partial\hat{d}_{s,n}}\frac{\partial\hat{d}_{s,n}}{\partial\hat{\mathbf{W}}^{*}_{n}}
\]
with $
\frac{\hat{d}_{s,n}}{\partial\hat{\mathbf{W}}^{*}_{n}}=\frac{1}{8}\Re\left\{\mathbf{h}^{a\T}_{s}\hat{\mathbf{W}}_{n}\mathbf{x}^{a}_{n}\right\}\mathbf{h}^{a}\mathbf{x}^{a\H}_{n}$. Moreover, for the special case of the metric function $\mathsf{d}(\cdot.\cdot)$ selected to be $
\mathsf{d}\left(\hat{d}_{s,n},d_{s,n}\right)=\left(\sqrt{\hat{d}_{s,n}}-\sqrt{d_{s,n}}\right)^{2}$, we have
\[
\nabla_{\hat{\mathbf{W}}^{*}_{n}}\mathsf{J}\left(\hat{\mathbf{W}}_{n}\right)\propto\left(\sum^{m}_{s=1}\left(\sqrt{\hat{d}_{s,n}}-\sqrt{d_{s,n}}\right)\mathbf{h}^{a}_{s}\right)\mathbf{x}^{a\H}_{n}.
\]
where from classical results on quaternion-valued adaptive filtering it follows that $\mu\in[0,(3\left\lceil\log_{2}\left(z\right)\right\rceil)^{-1})$ will yield convergent behaviour. Note that during training, $d_{s,n}$ is known and $\hat{d}_{s,n}$ is calculable from $\{\mathbf{x}_{n},\mathbf{h}_{s},\hat{\mathbf{W}}_{n}\}$ that are known parameters. Moreover,  $\hat{d}_{s,n}$ can be approximated from averaging measurements over multiple realisations. 

\section{Numerical Example}

\begin{figure}[h]
\centering
\includegraphics[width = 0.92\linewidth, trim = 0cm 0.2cm 0cm 1.4cm]{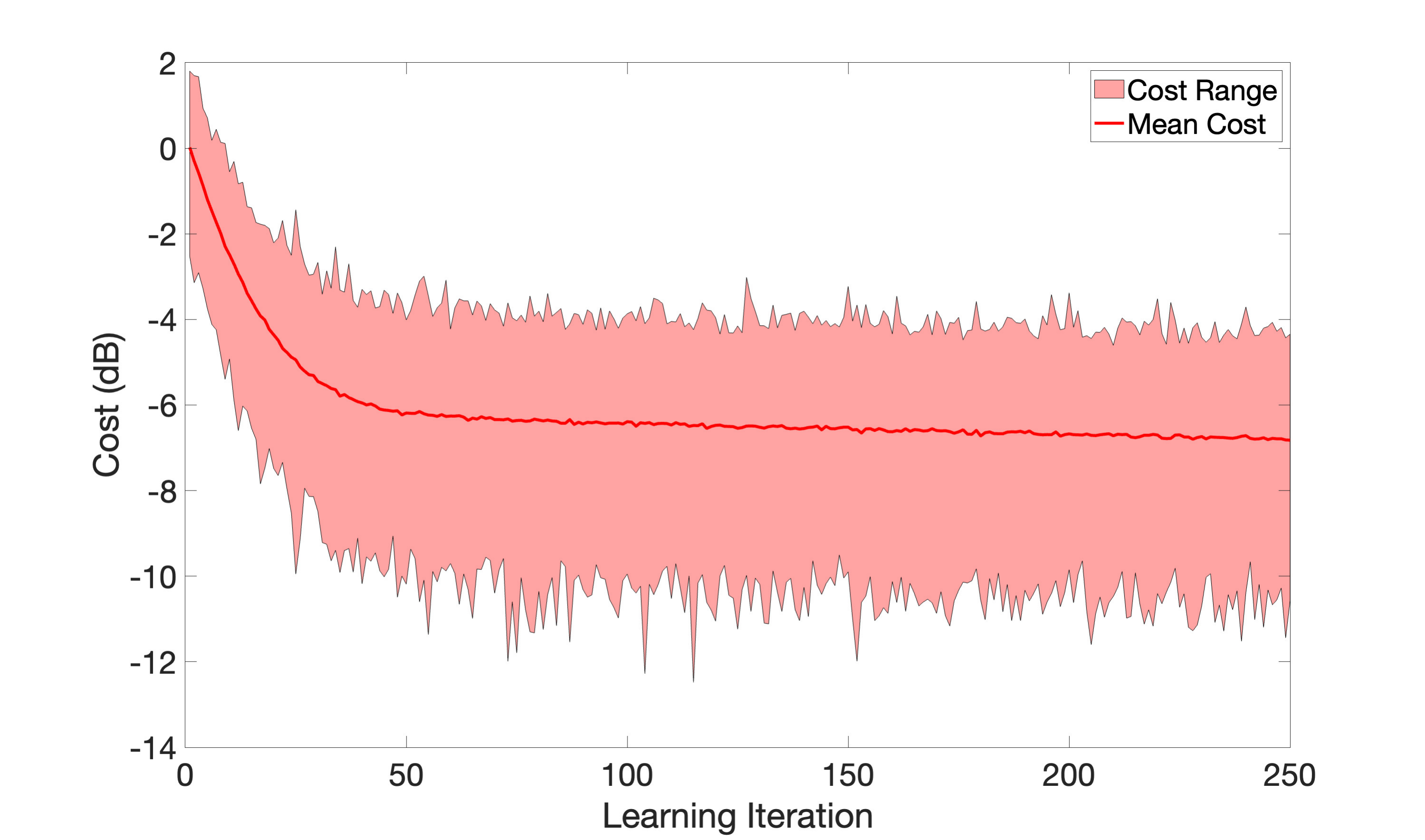}
\caption{The cost, $10\log\left(\frac{1}{8}\sum^{8}_{s=1}\mathsf{d}\left(\hat{d}_{s,n},d_{s,n}\right)\right)$, is shown for $10^{3}$ independent trials of the derived algorithm. The cost across all trials lie within the light red region, while the solid red line showing the average over all trials.}
\label{Fig:Error}
\end{figure}

In this section, the training performance of the proposed algorithm is demonstrated via a numerical example. To this end, an $8$-qubit system was considered, with observation available alongside the $\imath$, $\jmath$, and $\kappa$ vectors of each qubit. In addition, $\mathsf{P}\left(\zeta|\mathcal{D}_{n}\right)$ was assumed to be a uniform distribution on the positive orthant of an $8$-dimensional quaternion sphere, so that $\mathbf{y}_{n}$ resided on the surface $\{\|\mathbf{y}_{n}\|^{2}=1,\forall s: \Re\{\mathbf{h}^{a\T}_{s}\mathbf{y}^{a}_{n}\}\geq0\}$. The $\mathbf{y}_{n}$ samples were passed through a circuit to obtain $\mathbf{x}_{n}$. The derived algorithm was used to estimate a circuit relating the pair $\{\mathbf{y}_{n},\mathbf{x}_{n}\}$. Fig.~\ref{Fig:Error} shows the learning curve over $10^{3}$ independent realisations of the algorithm. Notice that the derived algorithms shows convergent behaviour. 

\begin{figure}[h]
\centering
\includegraphics[width = 0.92\linewidth, trim =  0cm 0.8cm 0cm 1.4cm ]{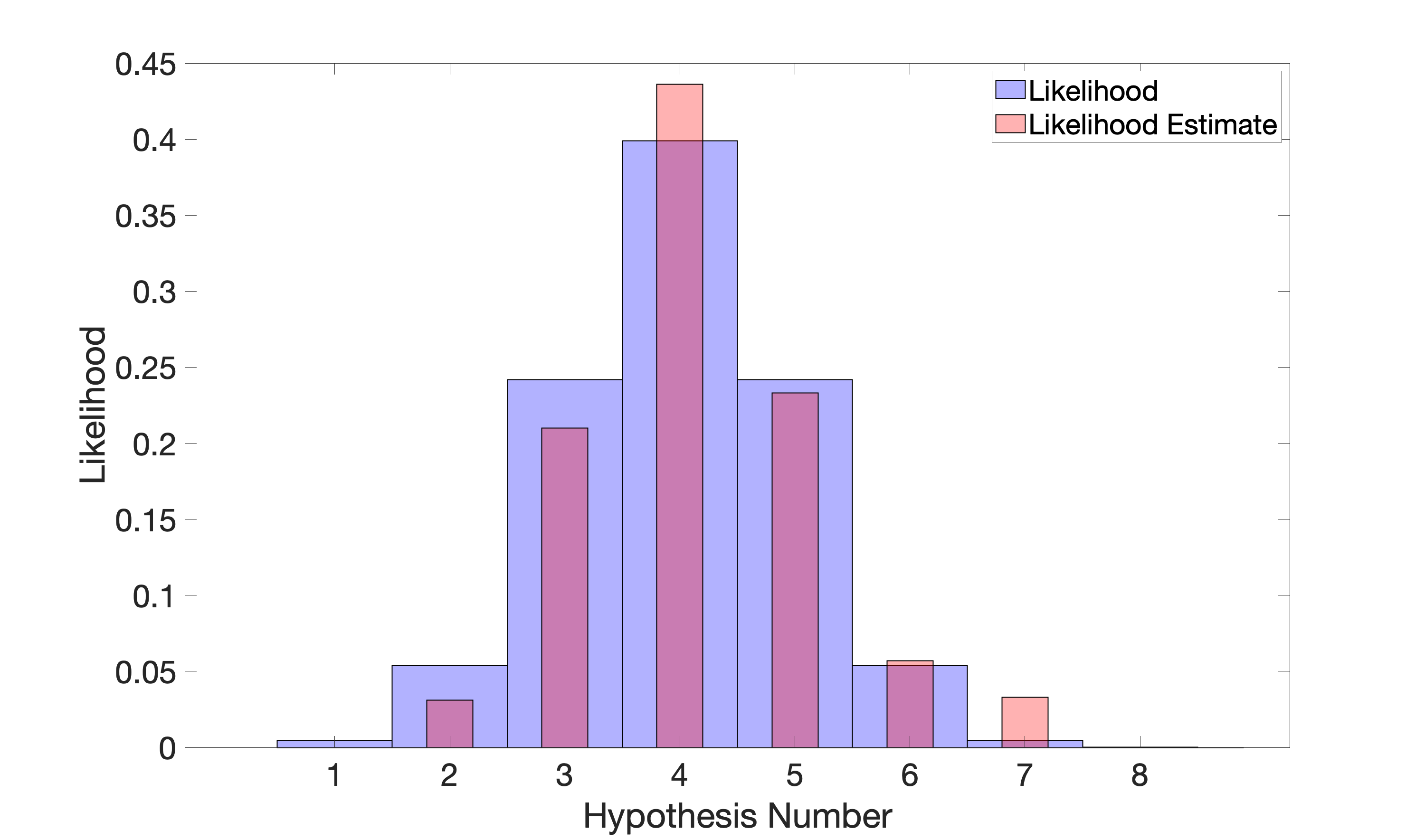}
\caption{Likelihoods of $\zeta'$ and its estimate.}
\label{Fig:Hist}
\end{figure}

In order to demonstrate operations of the estimated circuit data points were generated from $\mathsf{P}\left(\mathcal{D}_{n}|\zeta'\right)$ where $\zeta'$ were a selection of $8$ hypothesis from the region representing $\mathbf{y}_{n}$ in the previous experiment. The estimated circuit was used to find $\mathsf{P}\left(\zeta'|\mathcal{D}_{n}\right)$. The likelihood function of $\zeta'$ is shown against its estimate in Fig.~\ref{Fig:Hist}. Note that the estimated circuit closely estimated the correct likelihood for $\zeta'$.

\section{Conclusion}

A quaternion-valued formulation of the adaptive learning problem on quantum computational devices was considered, where a quaternion-valued model for implementing computation and measurement operation on qubits was presented and an adaptive learning algorithm was derived. The quaternion-valued framework allowed for use of the $\mathbb{HR}$-calculus to derive an intuitive and straightforward adaptive learning algorithm. In addition, the quaternion-valued formation was used so that recent advanced in quaternion-valued adaptive filtering can be leveraged to establish convergence and observability conditions. Performance of the resulting technique in learning relations between conditional probabilities was demonstrated in a simulation example. 








\bibliographystyle{IEEEtran}
\bibliography{ref}
%
%
%


\end{document}